\documentclass[apjl]{emulateapj}

\shorttitle{LONG TERM EVOLUTION OF PLANET-INDUCED VORTICES}
\shortauthors{Fu, Li, Lubow \& Li}

\begin{document}

\title 
{Long Term Evolution of Planet-Induced Vortices in Protoplanetary Disks}

\author{Wen Fu{$^{1,2}$}, Hui Li{$^2$}, Stephen Lubow{$^3$}, Shengtai Li{$^2$}}
\affil{{$^1$}Department of Physics and Astronomy, Rice University, Houston, TX 77005, USA; wf5@rice.edu\\
{$^2$}Los Alamos National Laboratory, Los Alamos, NM 87545, USA\\
{$^3$} Space Telescope Science Institute, Baltimore, MD 21218, USA}

\begin{abstract}

Recent observations of large-scale asymmetric features in protoplanetary 
disks suggest that large-scale vortices exist in such disks. 
Massive planets are known to be able to produce 
deep gaps in protoplanetary disks. The gap edges could become 
hydrodynamically unstable to the Rossby wave/vortex instability 
and form large-scale vortices. 
In this study we examine the long term evolution of these vortices by
carrying out high-resolution two dimensional hydrodynamic simulations 
that last more than $10^4$ orbits (measured at the planet's orbit). 
We find that the disk viscosity has a strong influence on both the emergence 
and lifetime of vortices. In the outer disk region where asymmetric features
are observed, our simulation results suggest that the disk viscous $\alpha$
needs to be low $\sim 10^{-5 }$ - $10^{-4}$ to sustain vortices to thousands 
and up to $10^{4}$ orbits in certain cases. The chance of finding a vortex feature in a disk then decreases with smaller planet
orbital radius.
For $\alpha \sim 10^{-3}$ or larger,
even planets with masses of 5 Jupiter-masses will have difficulty either producing or sustaining
vortices.  We have also studied the effects of different disk temperatures and planet masses. 
We discuss the implications of our findings on current and future 
protoplanetary disk observations.

\end{abstract}


\section{Introduction}

The new Atacama Large Millimeter/submillimeter Array (ALMA), although still 
at an early configuration, provides unprecedented resolution and sensitivity in the 
(sub)mm wavelength range. Recent high-fidelity 
ALMA images have revealed high contrast dust asymmetries in the outer parts of 
transitional disks around stars LkH$\alpha$ \citep{isella13}, Oph IRS 48 
\citep{marel13}, HD 142527 \citep{casassus13, fukagawa13}, 
SAO 206462 and SR 21 \citep{perez14}. These findings strongly
suggest that the disk dust (and presumably gas) distribution is non-axisymmetric and
the mechanisms for the generation and sustainment of these asymmetries need
explanation. 

Two scenarios have been suggested that could lead to strong asymmetries 
in protoplanetary disks, both of which use the excitation of Rossby wave instability 
\citep[RWI][]{lovelace99, li00}. 
The RWI is a global non-axisymmetric instability which can be 
excited when there is a strong gradient and an inflexion point in the disk's radial 
potential vorticity profile. Many of the previous studies of RWI, in terms of both
its linear stability properties and nonlinear
evolution, used an assumed and idealized structure (such as a bump or an edge) 
in the disk radial density profile
\citep{lovelace99, li00, meheut10, meheut12a, meheut12b, meheut12c, meheut13, richard13}. These structures, under appropriate conditions
(as detailed in the linear theory calculations in Li et al. 2000), will form non-axisymmetric unstable modes.
At the nonlinear stage, these non-axisymmetric modes develop into vortices and they
further merge to form one large vortex \citep[e.g.][]{li01}. 

To explain the observed features in the transition disks, one scenario envisions that
there exists sharp viscosity transitions in the radial direction of disks 
that may arise near the edges of dead zones. Sharp viscosity change, thus sharp density change at outer edge of a dead zone
can excite RWI \citep{regaly13, lyra12}. While the evolution of RWI
will try to smooth out the density variation, the continuous accretion from material
at large radii in the disk provides a driving effect on the RWI.

The other scenario is to explore 
the consequences of massive planets inside the disk \citep{lin86a, lin86b}, especially in connection
with the inner disk holes/cavities discovered in dozens of circumstellar disks 
\citep{forrest04, andrews11, espaillat10, kraus13, rosenfeld13, dodson11, zhu11, kraus12, dobinson13, ruge13}.
When a planet becomes massive enough, it carves out a deep gap around its orbit.
Due to its significant density variations and the corresponding angular velocity
adjustments, the gap edge can typically excite the RWI \citep[e.g.][]{li05}, 
leading to the formation of vortices. 
Further nonlinear development of the RWI can lead to the formation
of a ``banana-shaped" asymmetric density enhancement \citep[e.g.][]{li05} or one large vortex \citep[e.g.][]{lin11, lin12, lin14}. 
One important condition for exciting RWI by a planet is that the disk viscosity needs to be
sufficiently low, as it has been empirically studied by various groups 
\citep[e.g.][]{dvb07, li09, yu10, lin11}. This scenario has been proposed to explain the
ALMA observation of Oph IRS 48 \citep{marel13}, as well as
the disk gap/hole \citep{ataiee13}. 

These vortices are potentially very important because they can efficiently 
trap dust particles 
\citep[e.g.][]{barge95, johansen04, inaba06, rice06, meheut12b, pinilla12, zhu12, zhu14a, birnstiel13, lyra13}, 
which in turn can produce asymmetric features in disk dust emission and 
help promote potential planet formation.  
Even though many previous studies have shown the generation of strong 
vortices in the disk, their long-term evolution, especially their survival time 
under different disk conditions, was left unaddressed. Though the exact
lifetime of these vortices and/or asymmetric features is difficult to pin 
down observationally, the general expectation is that they need to 
survive up to $\sim$ disk lifetime at tens of AU distances. 

In this {\em Letter}, we present high-resolution, long-term two-dimensional
simulations of disk-planet interactions that span $> 10^4$ orbits at the location of
vortex and we have explored the effects of several key
disk/planet parameters on the vortex lifetime, including planet mass, disk viscosity, 
and disk temperature. In Section 2, we present the detailed set-up of our numerical simulations. 
We summarize our main results in Section 3, and discuss the implication of our results in Section 4.

\section{Numerical Setup}

In our study, the protoplanetary disks are assumed to be geometrically thin so that the 
hydrodynamical equations can be reduced to two-dimensional Navier-Stokes equations 
by considering vertically integrated quantities. We adopt an isothermal equation of state 
$P=c_s^2\Sigma$ where $P$ is the vertically integrated pressure, $\Sigma$ is the 
surface density and $c_s$ is sound speed. Simulations are 
carried out using our code \texttt{LA-COMPASS} (Los Alamos Compuational AStrophysics Suite). 

The planet is taken to reside on a fixed circular orbit at radius $r_{p}$ with 
Keplerian orbital frequency $\Omega_{p}$.  
We adopt dimensionless units in which the unit of length is $r_{p}$ and 
the unit of time is $1/\Omega_{p}$.
In dimensionless units, the disk is modeled between $0.2\le r \le 6.48$ 
with the planet at  $r=1$. We consider two mass ratios of the planet to the 
central star $\mu=M_{p}/M_{\star}=0.001$ and $0.005$, corresponding to a  
$1 M_{J}$ planet and a $5 M_{J}$ planet given a one solar mass central star. Planet mass is ramped up to its final value in the first
10 orbits. A smoothing length $r_{s}=0.6r_{H}$ is applied to the 
gravitational potential of the planet. We choose a power-law profile for both 
initial disk surface density and disk temperature of the form $\Sigma \propto r^{-1}$, 
$c_{s}\propto r^{-0.5}$. The disk aspect ratio given by $h/r = c_{s}/(\Omega r)$
is nearly independent of $r$ (hereafter we will use $h$ to stand for the dimensionless
disk temperature). The initial disk mass is about $1 M_{J}$.  
The dimensionless kinematic viscosity $\nu$ (normalized by $r_p^2 \Omega_p$) 
is taken to be spatially constant and ranges from $\nu=10^{-8}$ to $\nu=10^{-5}$. 
The Shakura-Sunyaev viscosity is related to $\nu$ by $\alpha=\nu/(\Omega h^2)$.   
All the simulations have a resolution of ($n_r \times n_{\phi}$) $=$ $3072\times3072$. 
The smallest Hill radius $r_{H}=0.07$ is thus resolved by 35 cells. We employ fixed value condition at boundaries. 
The initial disk surface density is completely smooth without an initial gap.
Our simulations typically last for $> 10^4$ orbits (at $r=1$). 

\section{Results}   
\begin{figure}[t]
\epsscale{1.0}
\plotone{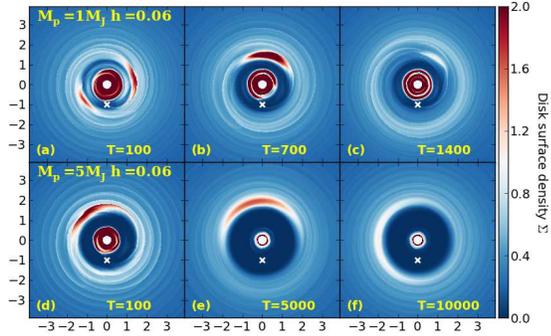}\\
\caption{Evolution of disk surface density showing how the vortex develops and evolves 
for two different planet masses, 
$M_{p}=1M_{J}$, $5M_{J}$. Both runs 
employ disk viscosity $\nu=1\times10^{-7} \, r_{p}^2 \Omega_{p}$. 
The location of the planet is marked by a white cross in every panel at dimensionless 
radius $r=1$.  Each row represents one simulation run with frames taken at different
time points in units of planet orbital period. (Color online)}
\label{fig:fig1}
\end{figure}

Figure~\ref{fig:fig1} shows the disk surface density evolution for two different planet masses.
For a $1 M_{\rm J}$ planet,  a gap in the disk can be developed quickly and the edges
of the gap become unstable, giving rise to vortices that quickly merge into a single vortex. 
This type of behavior is quite general for all massive planet cases we have studied.
This vortex can last for slightly more than $10^3$ orbits (Fig.~\ref{fig:fig1}(c)), then it finally disappears (when azimuthal density variation across the vortex falls below $\sim 10\%$ ).

For a $5 M_{\rm J}$ planet (Fig.~\ref{fig:fig1}(d-f)), a single vortex remains robust at 5000 orbits 
and persists even after $10^4$ orbits in Fig.~\ref{fig:fig1}(f). We see that after increasing the planet mass by a factor of 5, the vortex lifetime 
becomes almost 10 times longer for the same disk conditions. We expect the vortex 
survival time to increase with planet mass because a more massive planet is able to 
clear a deeper gap. The planet creates and maintains a sharper density jump at the 
gap edge that drives a stronger RWI. The vortex induced by the more massive 
planet covers a larger azimuthal range (see Fig.~\ref{fig:fig1}(b) and Fig.~\ref{fig:fig1}(e)).

\begin{figure}[t]
\epsscale{1.0}
\plotone{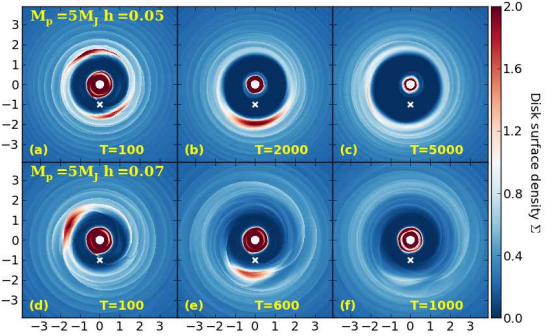}\\
\caption{Similar to Fig.~\ref{fig:fig1} except we now fix $M_{p}=5M_{J}$ and vary disk temperatures 
($c_{s}/(r\Omega)=h=0.05$, $0.07$). Again, both runs 
employ disk viscosity $\nu=1\times10^{-7} \, r_{p}^2 \Omega_{p}$. 
Together with Fig.~\ref{fig:fig1}(d)-(f), we see the vortex lifetime is not monotonic with the disk temperature. (Color online)}
\label{fig:fig2}
\end{figure}

Runs presented in  Fig.~\ref{fig:fig2} all have the same planet mass 
$M_{p}=5M_{J}$, but different disk temperatures, $c_s/\Omega|_{r=1} = h$. 
Both cases have vortex lifetime only on the order of a few thousands of orbits. Together with Fig.~\ref{fig:fig1}(d-f), 
$h=0.06$ seems to be the optimal disk temperature for the purpose of disk vortex survival time. In that case, the vortex lifetime is 
$\sim 13000$ orbits. We see that disk temperature has a very interesting nonmonotonic 
effect on the disk vortex lifetime. Either a higher or lower disk temperature results in 
more rapid vortex damping. 

\begin{figure}[t]
\epsscale{1.0}
\plotone{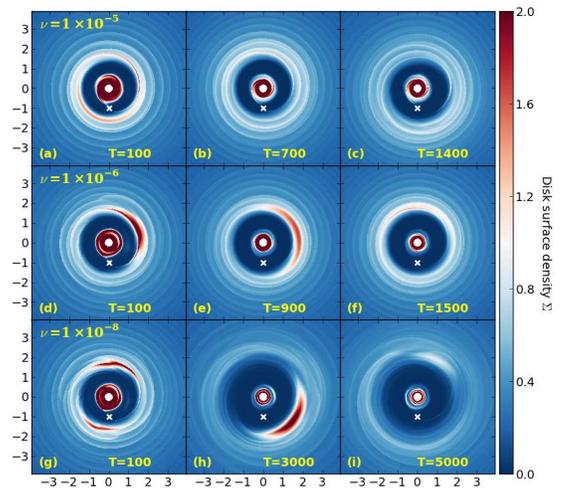}\\
\caption{Similar to Fig.~\ref{fig:fig1}, except that we fix the planet mass as $5M_{ J}$
and disk temperature $c_{s}/(r\Omega)=h=0.06$ but vary disk viscosities 
($\nu=10^{-5}$, $10^{-6}$, $10^{-8} 
\, r_{p}^2 \Omega_{p}$). Together with Fig.~\ref{fig:fig1}(d)-(f), we see the vortex lifetime is not monotonic with disk viscosity. (Color online)
}
\label{fig:fig3}
\end{figure}

\begin{figure}[t]
\epsscale{1.0}
\plotone{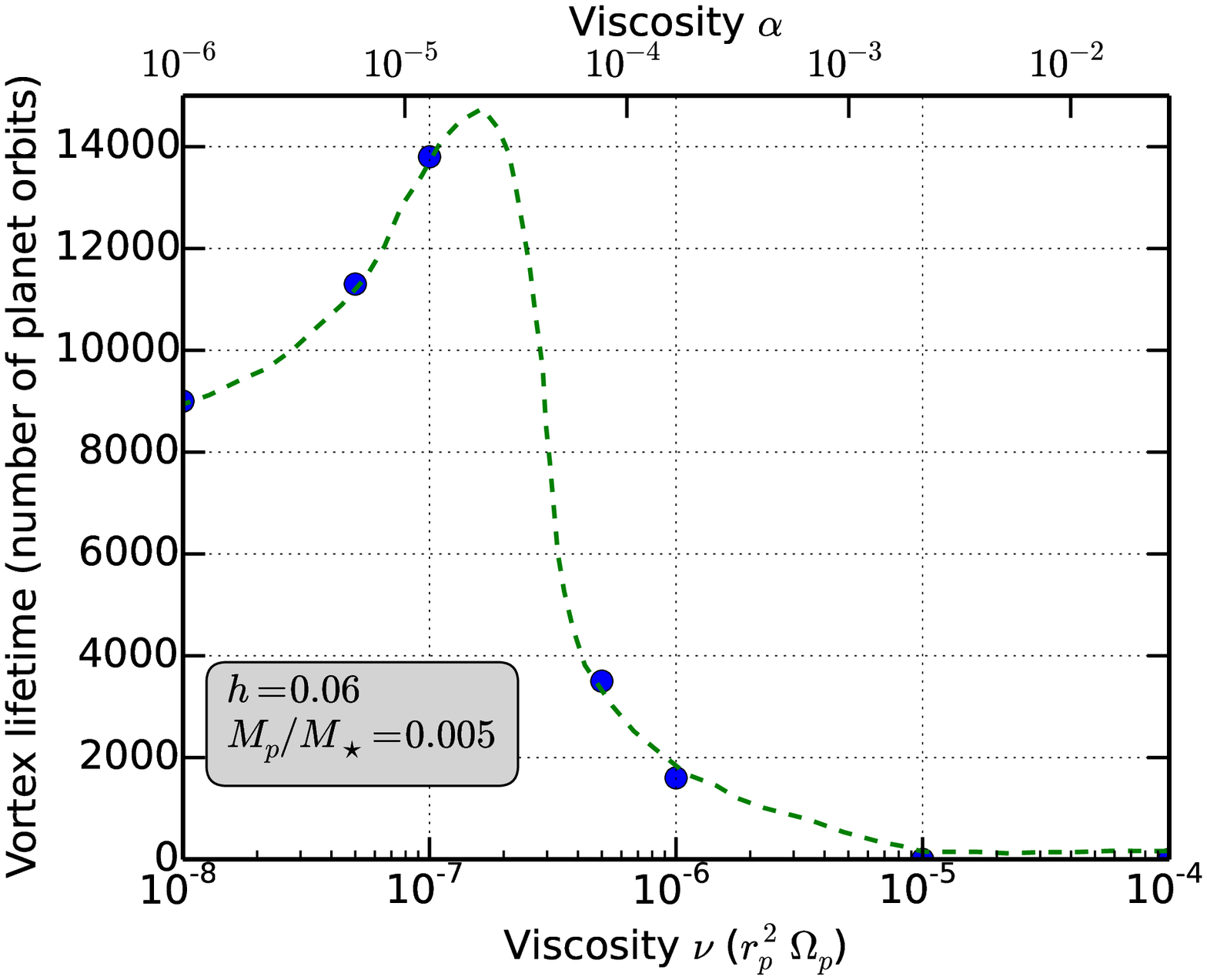}
\plotone{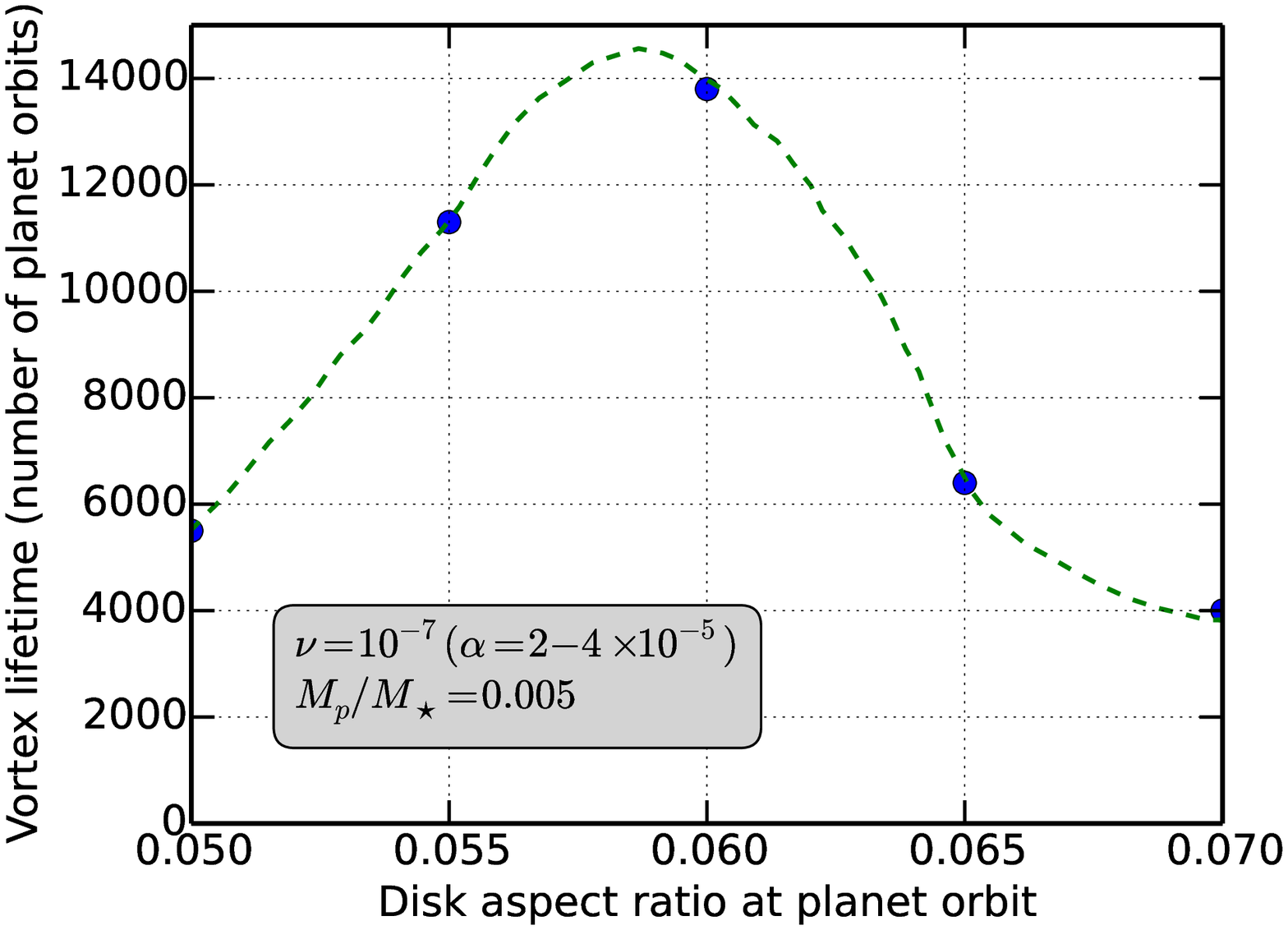}
\caption{Disk vortex lifetime (in units of number of planet orbits) as a function of viscosity at fixed $h=0.06 r_p$ (top) and temperature at fixed $\nu=1\times 10^{-7}$ (bottom). A vortex is deemed ``dead" after either the averaged azimuthal density variation or the averaged azimuthal potential vorticity variation within 10H (scale height) wide band around the vortex drops below 10\%.  The dashed lines are rough interpolations. }
\label{fig:fig4}
\end{figure}

A similar effect can also be seen in Fig.~\ref{fig:fig3} where we show runs with 
three additional disk viscosities ($\nu=10^{-5},\ 10^{-6},\ 10^{-8}$). 
Note that our code has numerical viscosity on the order of $10^{-9}$ or less. 
For $\nu=1\times10^{-5}$ (first row), the disk is barely able to form a discernible nonaxisymmetric feature even though there seems to be a clean gap. 
Any vortex disturbance gets damped out in a very short time (a few hundreds of orbits). 
For $\nu=1\times10^{-6}$ (second row), the vortex evolution is very similar to that 
for $\nu=1\times10^{-7}$ (Fig.~\ref{fig:fig1}(d-f)),
except that vortex lifetime is almost 10 times shorter. One would expect an even 
longer vortex lifetime for an even smaller viscosity because damping should decrease 
with smaller viscosity. Surprisingly, in the case of $\nu=1\times10^{-8}$ 
(third row), vortex lasts for significantly shorter time than in the case of 
$\nu=1\times10^{-7}$. Therefore disk viscosity affects vortex lifetime also in a 
nonmonotonic way. A viscosity value $\nu=1\times10^{-7}$ seems to be optimal for 
vortex survival with $M_{p}=5M_{\rm J}$ and $h=0.06$. Vortex suppression at large 
disk viscosity has been found before 
\citep{dvb07, li09, lin11, isella13, ataiee13}. But previous studies have only considered  
$\nu > 1\times10^{-7}$ and concluded the effect is monotonic. 
If viscosity is above some threshold ($\sim 10^{-5}$ in our runs), 
vortex formation can also be completely suppressed. 
The dependence of vortex lifetime on viscosity and temperature is summarized in Fig.~\ref{fig:fig4}, which includes more cases than we presented in Figs.~\ref{fig:fig1} and \ref{fig:fig2}. We will give a tentative explanation for this behavior in Section 4. 

\begin{figure}[t]
\epsscale{1.0}
\plotone{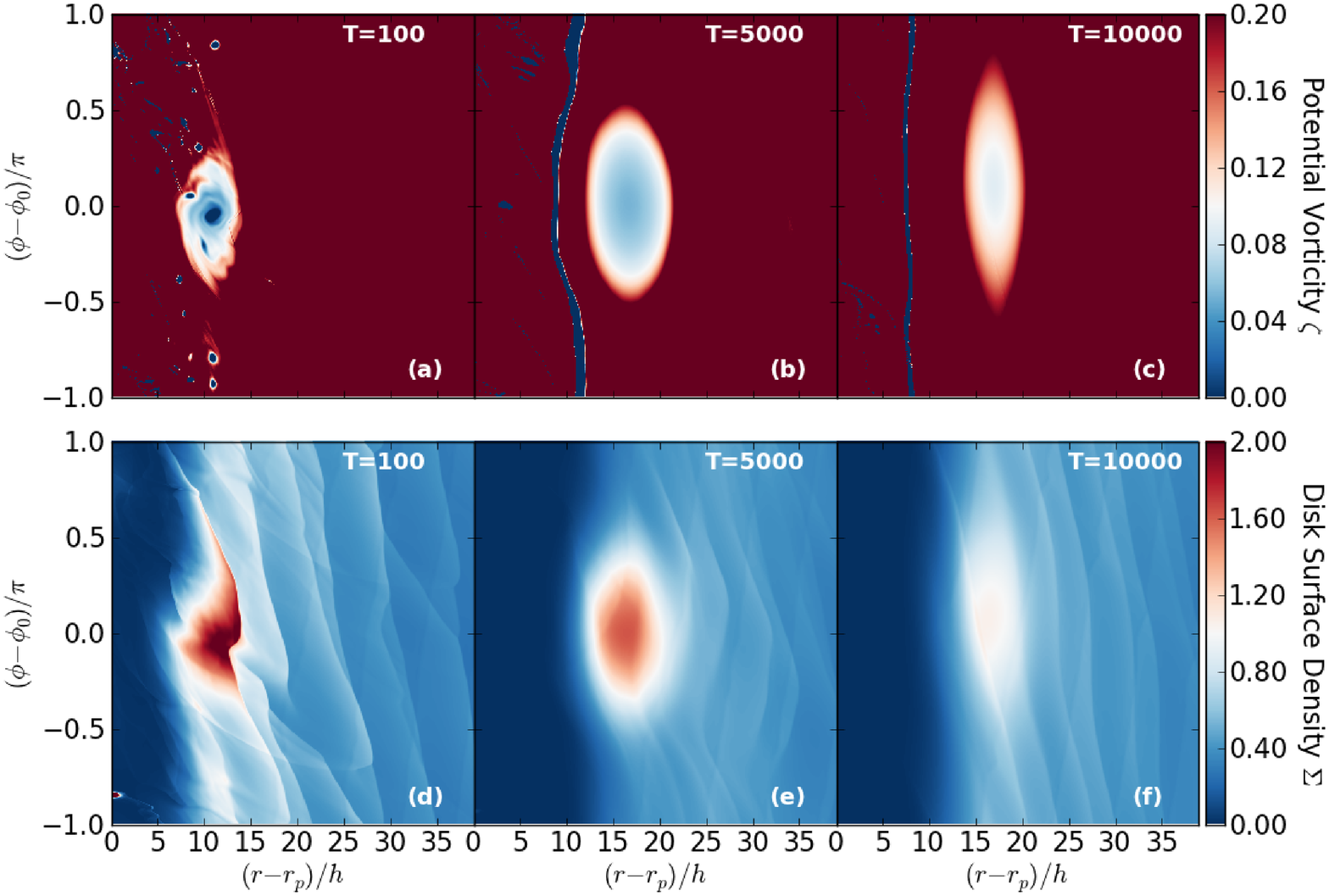}
\epsscale{1.0}
\plotone{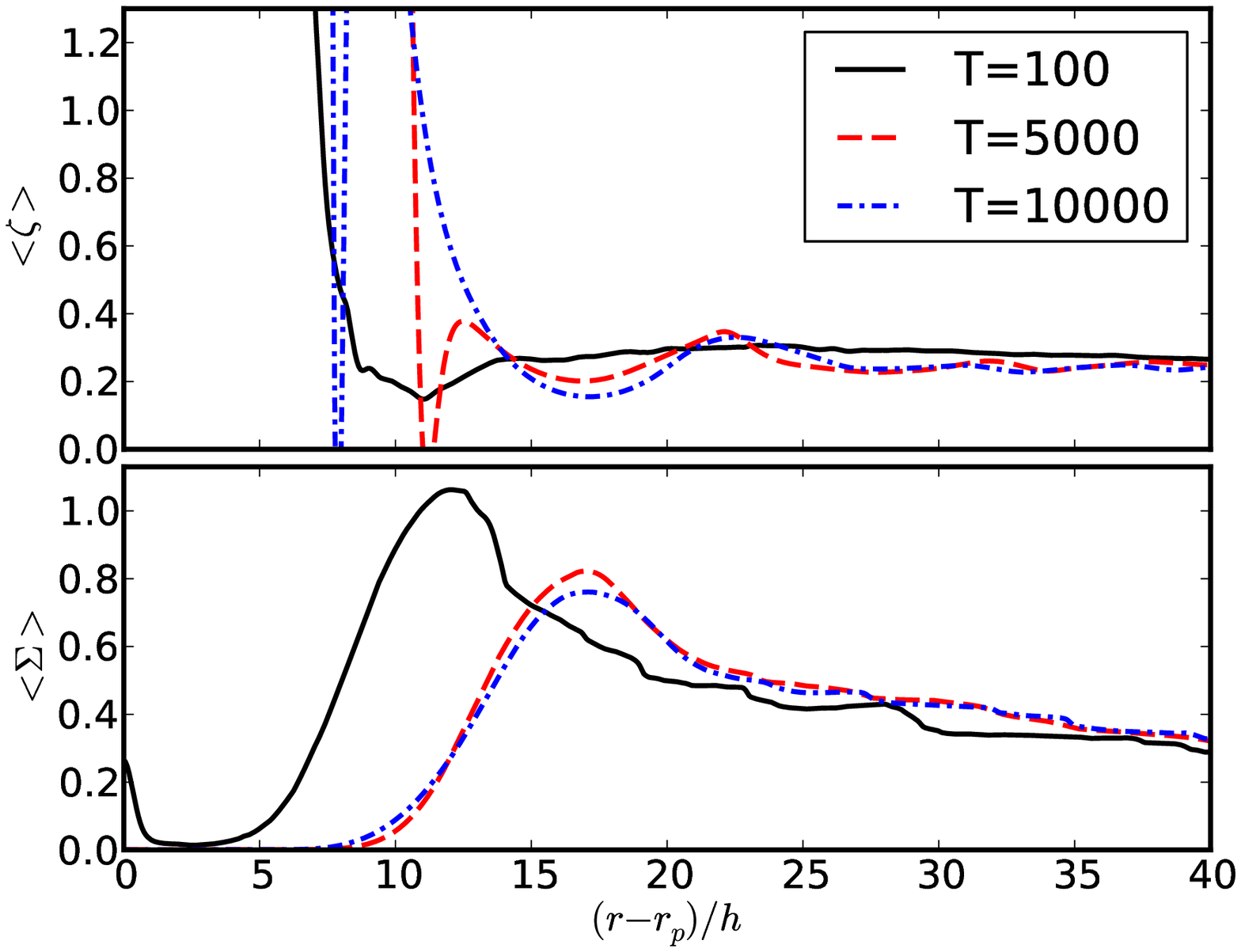}\\
\caption{
(Top) Color contours of disk potential vorticity and surface density distributions 
at different times for 
the run with planet mass $M_{p}=5 M_{J}$, viscosity $\nu=10^{-7} \, r_{p}^2 \Omega_{p}$, and dimensionless  disk temperature $c_{s}/(r\Omega)=h=0.06$.  
Time is in units of the planet orbital period. $\phi_{0}$ is the azimuthal coordinate of vortex center.  (Bottom) Azimuthally averaged disk potential vorticity and surface density profiles for the
same times as the upper plots. (Color online)
}
\label{fig:fig5}
\end{figure}

We now consider the evolution of vortex in more detail. 
The upper part of Fig.~\ref{fig:fig5} shows the evolution of $\zeta(r,\phi)$ and 
$\Sigma(r,\phi)$, where $\zeta = (\mathbf{\nabla} \times\mathbf{v})_z/\Sigma$ 
is the potential vorticity (PV). The vortex appears as 
a localized region of low PV (Fig.~\ref{fig:fig5}(a-c)) because the surface density
is higher in those regions. To ease comparison, we have shifted the plots azimuthally
so that the vortex is at the center in each Panel (a-c and d-f).  
Due to the large velocity perturbation and very low surface density, 
the PV within the gap region ($(r-r_{p})/h<10$) is much 
higher than in other regions of the disk. We set an upper cutoff on our color scale 
in order to make the vortex more clearly visible.  
The lower part of Figure~\ref{fig:fig5} shows the azimuthally averaged PV, 
defined as $\left< \zeta \right>=\left< (\mathbf{\nabla} \times\mathbf{v})_z/\Sigma \right>$, 
and disk surface density 
$\left< \Sigma \right>$ profiles at different times. 
As the surface density profile associated with the gap widens due to the continuous driving
by the planet (see the shift of in $\left< \Sigma \right>$ curves as a function of time),  the
minimum of PV profiles  $\left< \zeta \right>$ also shifts to larger $r$, moving from 
$(r - r_p)/h \sim 10$ at T$=100$ to $(r - r_p)/h \sim 17$ at T$= 5000 - 10000$. 
Correspondingly, the radial location of the vortex also moved the same amount. This is
quite consistent with the prediction of RWI theory where 
the disk vortex is formed where potential vorticity has a local minimum. 
In addition, from T=100 to T=10000, the vortex stretches azimuthally and 
its local strength is decreasing, as indicated by the increase of the PV at the
vortex center in going from Panels (a) to (c) of Fig.~\ref{fig:fig5}. 
The azimuthally averaged radial profile of PV between T=5000 and T=13500 
does not show much difference for $(r-r_p)/h > 12$. In fact, the PV minimum at 
T=10000 is even slightly deeper than the one at T=5000. At late time (T$\geq 10^{4}$), while both the disk 
surface density and average PV profiles have changed very small amount, the
vortex is getting narrower radially and is gradually being damped. 

\section{Discussion and Summary}

We have considered the interaction of gaseous protoplanetary disks with high mass 
planets that are in circular orbits with orbital frequency
$\Omega_p$ and radius $r_p$ from a central star.
We have sampled the parameter space of different planet masses, 
different disk viscosities, and different disk temperatures and investigated how 
these parameters affect the lifetimes of disk vortices. We find that higher planet 
mass generally leads to longer vortex lifetimes, given the same disk viscosity and 
temperature. This result occurs because a more massive planet carves out a 
cleaner gap and promotes a stronger RWI at the gap edge. Both disk viscosity and 
disk temperature have nonmonotonic effects on the vortex lifetime. We find that the 
optimal viscosity and dimensionless disk temperature
values for vortex longevity are $\nu = 10^{-7} \, r_p^2 \Omega_p$ ($\alpha \simeq 3\times 10^{-5})$ and 
$h = 0.06 r_p$.  Higher or lower values of $\nu$ or $h$ would either shorten the 
vortex lifetime or inhibit vortex formation (see Fig.~\ref{fig:fig4}). In all our runs, we do not see the ``return'' of the vortex after its
disappearance.

We do not believe that the lifetimes of the vortices in our simulations are 
determined by viscous damping because all our runs have vortex lifetimes that 
are significantly shorter than the viscous timescale on the vortex scale. 
In addition, we find that even lower viscosity (e.g., $\nu \sim 10^{-8}$) actually 
results in shortened vortex lifetimes. Instead, we speculate that the vortex is 
damped by shocks. Several competing effects are at play which jointly determine
the evolution and lifetime of the vortex. 
For lower viscosity or lower temperature, on one hand, 
a higher mass planet is able to create a sharper gap 
edge and thus form stronger disk vortices.
However, they also enhance spiral 
shocks that act to damp the vortex. 
For high viscosity or high temperature, shocks produced by the planet are weaker but 
the planet cannot create a sharp edge. Consequently, 
in this regime the planet either does not induce vortex formation at all or it only excites 
a weak vortex which damps quickly. Therefore,  intermediate values for disk viscosity and 
temperature provide the longest lived vortices that nearly
balance the driving with damping of disk vortex.  

Our findings can be compared with several recent observations. 
The longest vortex lifetime we find is $\sim 10^4$ orbits for $M_{p}=5M_{J}$, 
$\nu=1 \times 10^{-7}  \, r_p^2 \Omega_p$, and $h=0.06 r_p$. The vortex is located at 
$r = 2 r_p$. For the other sets of 
parameters that we have tried, the vortex lifetime spans from 0 to a few $\times10^{3}$ orbits. 
To explain relatively large disk gaps with dust emission asymmetries
\citep{casassus13, fukagawa13}, the planet needs to be located far
from the central star. For a slightly smaller hole as seen in Oph IRS 48,
if we take the planet to be located at a radius of 20 AU, then the vortex is located at about
40 AU, and the disk inner gap has radius $\sim$ 35 AU. Then the longest vortex lifetime
we simulated implies that this vortex can live for up to $10^6$ yrs (assuming that the
central star is one solar mass). Interestingly, 
these conditions are very similar to those found in the recent ALMA images of the 
disks around star Oph IRS 48 \citep{marel13} and stars SAO 206464, 
SR 21 \citep{perez14}.

The Rossby wave/vortex instability is due to the gas. 
But its interaction with the dust leaves observable signatures as seen by ALMA.
To better connect with observations such as ALMA images, one needs to also model 
the dust dynamics and radiative transfer. It has been suggested that 
dust can remain concentrated after the gas vortex has decayed  
\citep{birnstiel13}. We assume
that gas and dust vortex lifetimes are roughly equal. Our preliminary
results (not described here) suggest they are within 30\% of each other.
The disk vortex  
needs to last for $\sim 10^6$ yrs in order to be responsible for the dust emission 
asymmetry in  ALMA images of transition disks. As we have shown before, the dust asymmetry at 
$\sim$ 40 AU can survive for $10^{6}$ yrs, but in a very small range of 
our simulation parameters. For dust asymmetry at $\sim$ 100 AU, 
the required number of orbits reduces to a few $\times 10^{3}$ orbits that can
 be realized for a broader range of disk parameters.
Due to the resolution of current ALMA configuration, all of the dust asymmetries 
are found to be far from the central star. With the most extended ALMA configuration, 
future observations will be able to resolve disk feature on scales closer
to the central star. On-going exoplanet surveys \citep{brandt14} could also shed light on the direct detection of forming 
massive planets in disks with these asymmetry features.

At a large orbital radius ($\sim 50$ AU),  the very low disk viscosity value ($\nu=10^{-7} \, r_p^2 \Omega_p$ or $\alpha=10^{-4}$ at $r = 2$ where the vortex is located)  
for vortex longevity implies that this region does not evolve viscously over the disk lifetime.
Such a low viscosity requires some explanation.
In the T Tauri phase, the observationally inferred accretion rates onto the central star suggest that $\alpha \sim 10^{-2}$ \citep{hartmann98}.
At that level of turbulence, we do not expect that vortices can form.
The magneto-rotational instability (MRI) is a likely source of turbulence
in the outer regions of a protoplanetary disk \citep{balbus91}.
MRI typically results in an $\alpha$ value $\alpha \ga 0.01$ that is again
 too high to permit the development of a vortex.
On the other hand, the efficiency of MRI is weakened considerably in certain regions of protostellar disks
due to nonideal MHD effects that result from the low levels of ionization \citep[e.g.][]{bai11}.
Recent simulations by \cite{zhu14b} indicate that sufficiently low levels of $\alpha$ and long vortex
lifetimes can be achieved through the nonideal effects of ambipolar diffusion.
 The reconciliation of the low viscosity requirements of vortex generation with the high viscosity requirement
 of accretion is unclear, possibly involving alternate accretion mechanisms. 

The results presented here represent some preliminary steps in trying to understand the 
joint evolution of planet, disk accretion, vortices and dust asymmetries in the outer parts of
the protoplanetary disk. Dust-gas interaction, disk self-gravity, more sophisticated viscosity profile, 3D structure could all affect disk vortex evolution to some extent. We plan to address these issues in future studies.

\section*{Acknowledgements}
Simulations in this work were performed using the Institutional Computing Facilities 
at LANL. WF, HL and SL gratefully acknowledge the support  by the LDRD and IGPP 
programs and DOE/Office of Fusion Energy Science through CMSO at LANL.  
WF and SL acknowledge support from NASA grant NNX11AK61G. We thank Til Birnstiel 
and Zhaohuan Zhu for valuable comments.

\end{document}